\documentclass[a4paper]{iopart}

\newcommand{\solm}{M_{\odot}}
\newcommand{\teff}{$T_{\rm{eff}}$}

\usepackage{harvard}
\usepackage{graphicx}
\usepackage{subfig}
\usepackage{longtable}

\begin{document}
\bibliographystyle{jphysicsB}

\title{The Age of White Dwarf Companions}

\author{S. Weston and R. Napiwotzki}

\address{Centre for Astrophysics Research, %
University of Hertfordshire, College Lane, Hatfield, Herts, AL10 9AB, UK}
\ead{s.1.weston@herts.ac.uk}

\begin{abstract}
We carried out a spectroscopic investigation of single lined white
dwarfs (WDs) in double degenerate (DD) systems and discuss their
binary evolution. Simulated spectra of the H$\alpha$ region are used
to derive upper limits on the temperature of the invisible component
and thus lower limits on the cooling age. This is done for a range of
hypothetical secondary masses and a minimum cooling age
deduced. Results are compared with the well known parameters of the
visible primary, which allows us to determine a lower limit for the
cooling age difference of both WDs. Most of the ten systems in our
sample have a minimum age difference of not larger than 0.5Gyr and
their small orbital separation is highly suggestive of at least one
unstable mass transfer phase. However, a stable first mass transfer
phase is feasible as the age difference is less then 1Gyr. The results
imply that unstable mass transfer is the most likely final contact
binary scenario to have occurred in DD systems but the first mass
transfer phase is not constrained.

\end{abstract}

\section{Introduction}

Between 5 and 10\% of white dwarfs (WD) reside in close binary systems
with a WD companion \cite{nap03,Mar00}, known as double degenerates
(DDs). DDs must have come into contact and fallen towards each other
as it is not possible for the stars to have evolved so close during
giant branch phases. Orbital periods range from hours to a few days
corresponding to seperations of 2-3$R_{\odot}$ at most. This implies
that the binary must have gone through phases of intensive interaction
affecting the evolution of the WD progenitors. One of the binary
components evolving up the giant branch (RGB or AGB) will fill its
Roche lobe and start to transfer mass to its companion. This can
result in three possible outcomes \cite{nel01}:
\begin{itemize}
\item Conservative/stable mass transfer –-- The transferred mass is
  accreted by the companion and no mass is lost to the
  surroundings. The size of the orbit remains relatively large and can
  even increase.
\item Standard common envelope –-- The rate of mass transfer is so
  high that not all of it can be accreted. Most of it is lost and
  forms a common envelope around the system. Friction causes a large
  reduction in the orbital seperation.
\item Envelope eject or double spiral-in –-- During the ejection of
  the common envelope both star’s lose their envelope. Orbital radius
  greatly reduced and age difference very small (within 0.1Myr).
\end{itemize}
Different formation time scales can be expected for the three
scenarios. Only very small age differences are allowed for DDs
resulting from the double spiral-in channel. Stable mass transfer
results in a substantial mass increase of the companion, which will
speed up its subsequent evolution. Since it is a very short lived
phase, no such effect is expected during the common envelope. We
derive lower limits on the age of the secondary WD and therefore an
age difference estimate of the primary and secondary's formation. This
may be used with the theoretical predictions to rule out possible
evolutionary scenarios for a binary system. The aim of this
investigation is to compare the resulting limits with predicted
distributions for the different channels.

\begin{figure}[!t]
       \includegraphics[width=2.5in,bb=30 165 565 690]{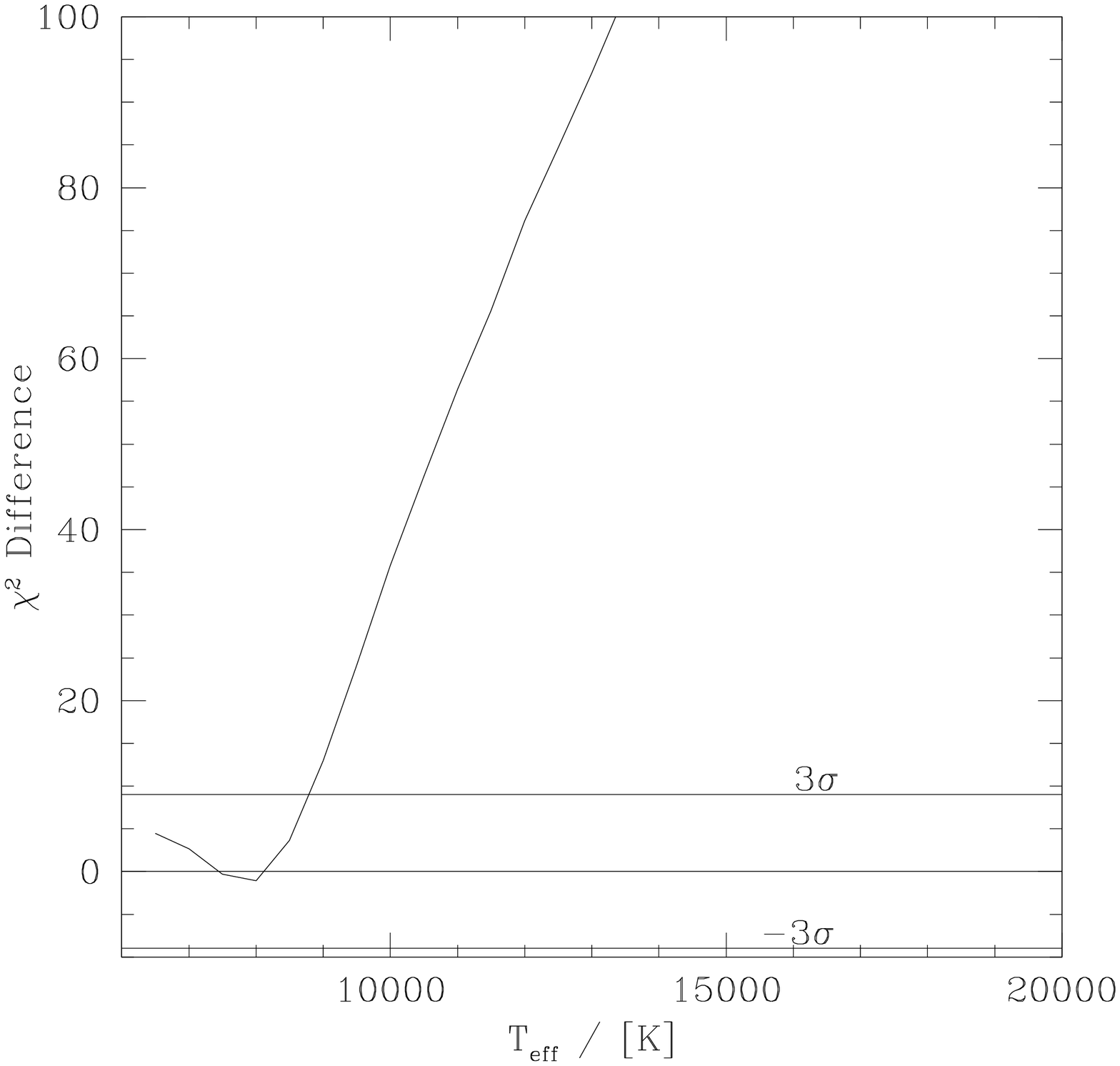}
    \hfill 
       \includegraphics[width=2.5in,bb=30 165 565 690]{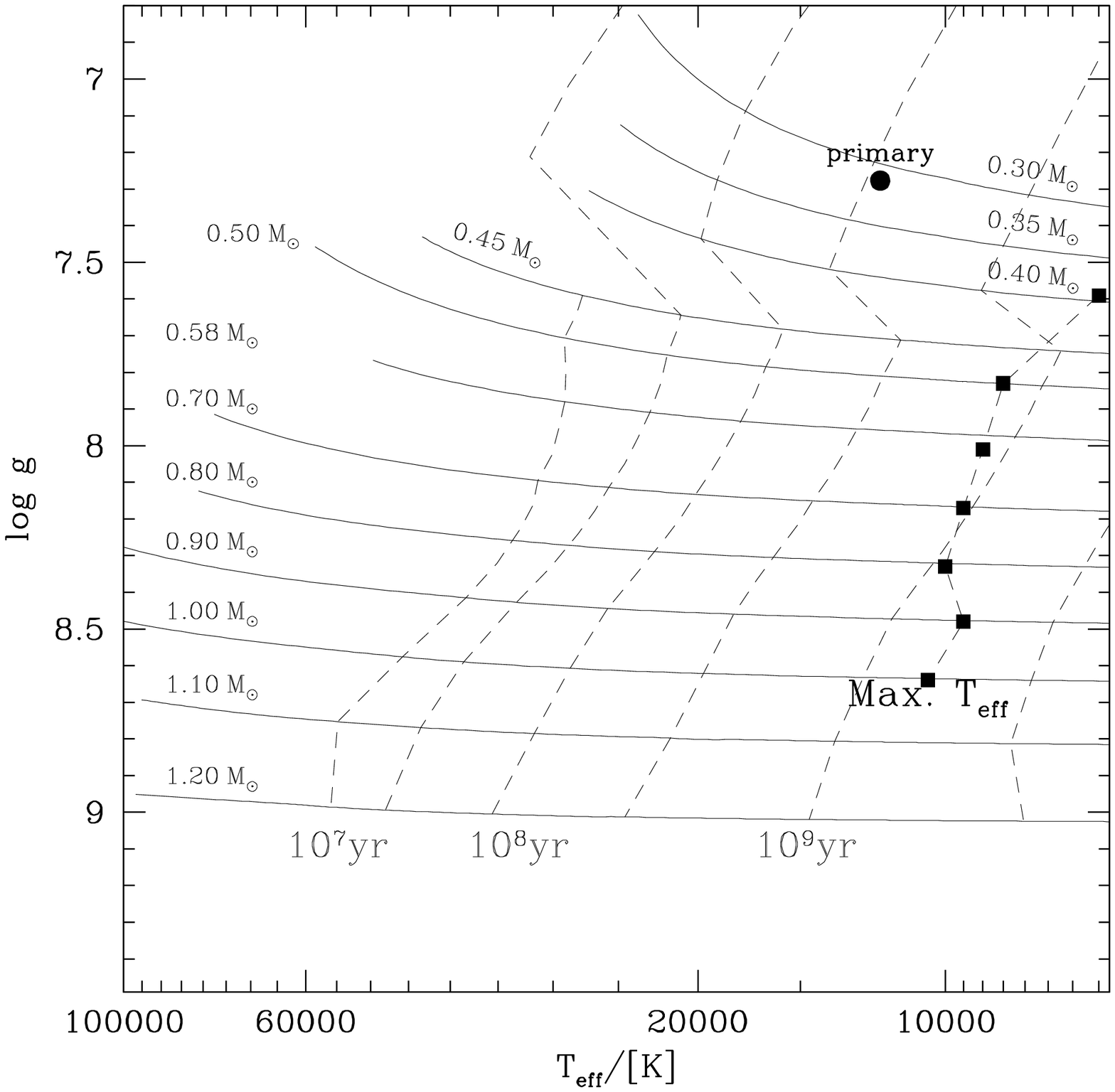}
    \caption{\small \textit{Left:} Fit quality as a function of
      temperature for a 0.5$\solm$ companion of HE0320$-$1917. The
      maximum possible temperature of the secondary is
      8500K. \textit{Right:} H-R diagram with maximum \teff\ for
      hypothetical seconday masses of HE0320$-$1917. Evolutionary
      tracks of \citeasnoun{ben99} for masses (solid lines) are shown
      along with cooling ages (dashed lines).}
    \label{fig:he0320}     
\end{figure}

\section{Determining Maximum Secondary Temperatures}

The DDs investigated have been classified SB1 in the Supernova type Ia
Progenitor surveY (SPY; Napiwotzki et al., 2003\nocite{nap03}),
meaning that no obvious contribution from the secondary is present in
the spectra. We use SPY spectra of the H$\alpha$ region to constrain the
secondary's contribution in a quantitative way. We consider just the
H$\alpha$ line because its core is the most prominent feature in the
spectra of DA WDs. SPY is a high resolution, high signal-noise
spectroscopic study which uses a large input sample of WDs combined
from many surveys. Systems are confirmed DD due to radial velocity
(RV) fluctuations. The data set contains six binary systems for which
orbital solutions and parameters for the brighter primary star are
already known and published in \citeasnoun{nel05}. Four other
non-published systems were used with well established primary
parameters. For the latter four systems the RV and thus position of
the H$\alpha$ core is not constrained. In these cases a worst case was
assumed with the secondary line core close to the primary's. Using a
fitting program, FITSB2 \cite{nap04}, synthetic spectra are created
and compared to the observed, and the $\chi^{2}$ calculated. The
parameters which produce fits outside 3$\sigma$ confidence limits can
be excluded with our attention focused on the possible solutions. The
last good fit defines a maximum temperature for a given mass component
of the secondary star. This was done for secondary masses of
0.4--1.0$\solm$ at a 0.1$\solm$ increment. Fig.~\ref{fig:he0320} shows
HE0320$-$1917 as an example.

\begin{table}[!t]
\caption{\small{Overview of the primary parameters and secondary
    limits. The columns give primary mass ($M_{1}$), effective
    temperature of the primary (T$_{\rm{eff,1}}$), cooling age of the
    primary ($\tau_{\rm{cool,1}}$), assumed secondary mass ($M_{2}$),
    maximum secondary temperature (Max. T$_{\rm{eff,2}}$) and minimum
    secondary cooling age (Min. $\tau_{\rm{cool,2}}$). The last column
    indicates if the fit quality increased by more than 3$\sigma$ with
    a secondary component, therefore making the system a SB2
    candidate. We show minimum ages for all hypothetical masses of
    HE0320$-$1917 and then the youngest calculated age with its
    corresponding mass for all other systems.}} \label{data_results}
\centering
\begin{tabular}{lrrlrrlr}

\hline \hline \small Object & $M_{1}$ & T$_{\rm{eff,1}}$ &
$\tau_{\rm{cool,1}}$ & $M_{2}$ & Max. &
Min. & SB2? \\
\small& & & & & T$_{\rm{eff,2}}$ & $\tau_{\rm{cool,2}}$ & \\
\small& [$\solm$] & [K] & [Gyr] & [$\solm$] & [K] & [Gyr] & \\

\hline \hline
\small
HE0320$-$1917 & 0.29 & 12 000 & 0.3327
& 0.4 & 6 500 & 2.337 & Y\\
& & & & 0.5 & 8 500 & 0.7505 \\
& & & & 0.6 & 9 000 & 0.8344 \\
& & & & 0.7 & 9 500 & 0.9409 \\
& & & & 0.8 & 10 000 & 1.116 \\
& & & & 0.9 & 9 500 & 1.685 \\
& & & & 1.0 & 10 500 & 1.561 \\

HE1511$-$0448 & 0.48 & 50 000 & 0.0001
& 0.4 & 41 000 & 0.0026 & Y\\

WD0326$-$273 & 0.51 & 9 300 & 0.6883
& 0.5 & 6 500 & 1.474 & -\\

WD1013$-$010 & 0.44 & 8 000 & 0.8449
& 0.5 & 6 500 & 1.474 & -\\

WD1210+140 & 0.29 & 32 000 & 0.0232
& 0.6 & 9 000 & 0.834 & -\\


WD0101$+$048 & 0.48 & 8 600 & 1.517
& 0.5 & 6 500 & 1.474 & -\\

HE0131$+$0149 & 0.33 & 14 500 & 0.2541
& 0.8 & 26 000 & 0.0646 & Y\\

WD0216$+$143 & 0.34 & 27 000 & 0.0433
& 0.5 & 30 000 & 0.0088 & -\\

WD0341$+$021 & 0.30 & 21 500 & 0.0747
& 1.0 & 26 000 & 0.1419 & Y\\

WD1824$+$040 & 0.31 & 14 000 & 0.2817
& 0.5 & 10 000 & 0.4492 & Y\\
\hline
\end{tabular}
\end{table}

\section{Discussion on Companion Ages and Possible SB2 Systems} 

\begin{figure}
  \includegraphics[bb=0 0 500 515,width=2.5in,height=2.5in,clip=true]{T6500_poster.ps}
  \hfill
\includegraphics[width=2.5in,bb=18 143 579 722]{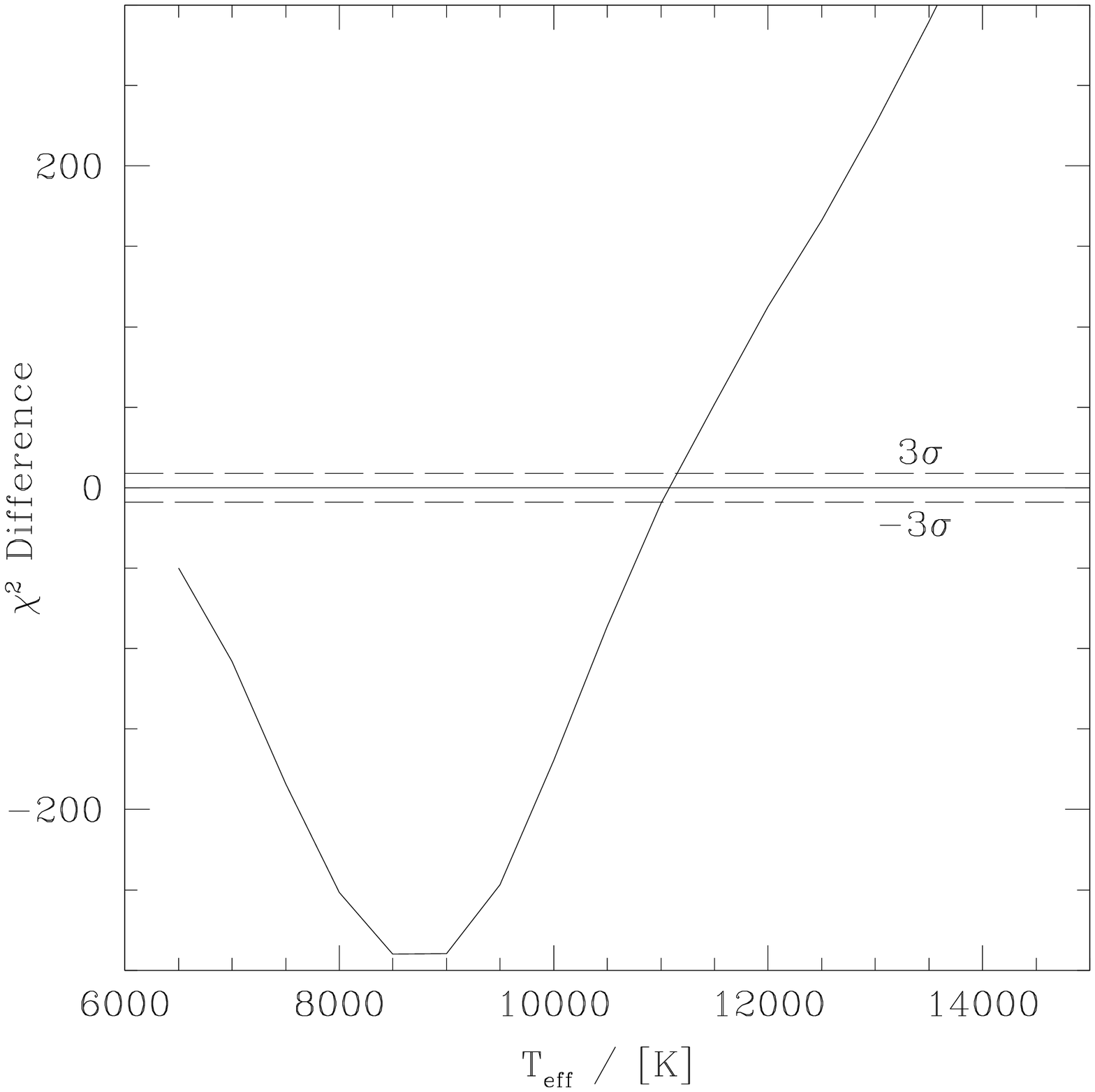}
    \caption{\small \textit{Left:} Fit to one of the spectra of
      WD1824+040 with an apparent detection of the seconqdary. The
      primary (1) and secondary (2) atmospheric parameters are
      displayed in the plot.  \textit{Right:} Summary of fit quality
      for various different temperature secondaries. The maximum
      temperature of the secondary is 11,000K.}
    \label{fig:wd1824}     
\end{figure}

The results are summarised in Table \ref{data_results} and suggest
nearly all of the systems companions have a larger cooling time then
the primary. Only in three systems the secondary could be younger but
it must be emphasised that we only determine lower age limits. In
addition, for one of these cases (and some other) we are constrained
by our cool model grid boundary of 6500K. In the cases where the
secondary is definitely older, all of the secondaries are older by more
than 0.1Myr therefore ruling out a double spiral-in
scenario. \citeasnoun{nel01} consider 1Gyr an approximate maximum age
difference for a secondary to have undergone an initial stable mass
transfer before a common envelope. In all but one of our sample this
is a feasible evolutionary path.\\ Some of our $\chi^{2}$ results
indicate possible detections of secondary components in the spectra;
HE0320$-$1917, HE1511$-$0448, HE0131+0149, WD0341+021 and WD1824+040
are all candidate SB2s and further analysis of the spectra may confirm
the features of the secondary.  Fig~\ref{fig:wd1824} shows the spectra
and $\chi^{2}$ plots of WD1824+040, a SB2 candidate. These systems may
be useful as they are close to the SB1/SB2 borderline and will show at
what mass and temperature one might expect to see features appear in
the combined spectra.

\section{Conclusion}
We test possible evolutionary scenarios of DDs by computing the
cooling age of ten systems and comparing the primary with hypothetical
secondary masses. The secondary age in nearly every case is larger
than the primary as expected. None of the older secondaries have a
small enough cooling age difference for a double spiral-in to have
occured. In all systems the minimum age difference is below 1Gyr,
therefore the first mass transfer phase may have been stable or
unstable. The results agree with current models which suggest, if the
hypothetical mass is close to the primary's the cooling ages are
similar. Additionally, the results of the systems without orbital
solutions show that limits can be obtained without knowing orbital
parameters of the system and with a conservative estimate a meaningful
limit can be deduced.

\section*{References}
\bibliography{wd_age_barca_poster}

\end{document}